\documentclass[10pt, conference]{IEEEtran}
\IEEEoverridecommandlockouts
\usepackage{cite}
\usepackage{amsmath,amssymb,amsfonts}
\usepackage{algorithmic}
\usepackage[pdftex]{graphicx}
\usepackage{graphicx}
\usepackage{textcomp}
\usepackage[table]{xcolor}
\usepackage{braket}
\usepackage{makecell}

\usepackage{comment}
\usepackage{booktabs}
\usepackage{here}
\usepackage{subfigure}
\usepackage{physics}
\usepackage{ascmac}
\usepackage[hyphens]{url}
\usepackage{multirow}
\usepackage{hyperref}
\usepackage{xspace}
\usepackage{framed}
\usepackage{inconsolata}
\usepackage{threeparttable}
\usepackage{pifont}
\usepackage{circledsteps}
\usepackage{listings}
\usepackage[normalem]{ulem}
\usepackage{tcolorbox}
\usepackage{enumitem}
\usepackage{breakurl}
\usepackage{caption}

\def\eg{e.g.\xspace}

\definecolor{diffstart}{named}{gray}
\definecolor{diffincl}{rgb}{0.0, 0.5, 0.0} 
\definecolor{diffrem}{named}{purple}

\lstdefinelanguage{diff}{
language=python,
basicstyle=\ttfamily\small,
columns=fullflexible,
float,
floatplacement=t,
frame=tb,
morecomment=[f][\color{diffrem}]{-}, 
morecomment=[f][\color{diffincl}]{+}, 
}

\newcommand{\rqone}{How well can Small Language Models fix bugs?}
\newcommand{\rqtwo}{How does quantization affect bug fixes?}

\begin{document}

\title{How Small is Enough? Empirical Evidence of Quantized Small Language Models for Automated Program Repair
}

\author{
\IEEEauthorblockN{
    Kazuki Kusama\IEEEauthorrefmark{1},
    Honglin Shu\IEEEauthorrefmark{1},
    Masanari Kondo\IEEEauthorrefmark{1},
    Yasutaka Kamei\IEEEauthorrefmark{1},
}
\IEEEauthorblockA{
    \IEEEauthorrefmark{1}Kyushu University, Fukuoka, Japan \\
}
}


\maketitle
\IEEEpubid{\begin{minipage}[t]{\textwidth}\ \\[10pt]
        \centering\normalsize{979-8-3315-9147-2/25/\$31.00~\copyright~2025 IEEE}
\end{minipage}}

\begin{abstract}
\textbf{Background:} Large language models (LLMs) have greatly improved the accuracy of automated program repair (APR) methods. However, LLMs are constrained by high computational resource requirements.  
\textbf{Aims:} We focus on small language models (SLMs), which perform well even with limited computational resources compared to LLMs. We aim to evaluate whether SLMs can achieve competitive performance in APR tasks.  
\textbf{Method:} We conducted experiments on the QuixBugs benchmark to compare the bug-fixing accuracy of SLMs and LLMs. We also analyzed the impact of int8 quantization on APR performance.  
\textbf{Results:} The latest SLMs can fix bugs as accurately as—or even more accurately than—LLMs. Also, int8 quantization had minimal effect on APR accuracy while significantly reducing memory requirements.  
\textbf{Conclusions:} SLMs present a viable alternative to LLMs for APR, offering competitive accuracy with lower computational costs, and quantization can further enhance their efficiency without compromising effectiveness.
\end{abstract}


\section{Introduction} \label{intro}

Debugging software systems, a costly process in development~\cite{britton2013reversible}, must be performed efficiently to reduce costs while identifying and fixing bugs.
To that end, \textit{Automated Program Repair (APR)} has been widely studied as a technique to automate the debugging process~\cite{Shariffeen2021TOSEM,Ye2021ESE}.
APR can automatically identify program bugs and generate patches to fix them~\cite{Weimer2009ICSE}.

Since 2022, \textit{Large Language Model (LLM)-based APR} has emerged~\cite{Xia2023ICSE}, achieving significantly higher accuracy than existing APR methods~\cite{Prenner2022ICSE,Fan2023ICSE}. 
Unlike traditional APR methods trained on a limited number of datasets, LLMs are trained on vast corpora containing up to hundreds of billions of text and code tokens, enabling them to fix even complex bugs.



Despite the high accuracy of LLM-based APR, it requires substantial computational resources~\cite{Yang2024robustness}.
For example, Code Llama-70B, which demonstrated high accuracy in APR~\cite{Huang2025ICSE}, contains 70 billion parameters and requires at least two NVIDIA A100 80GB GPUs for inference. Such a requirement makes it impractical for developers to use these models in their standard development environments, such as on personal laptops.
Since debugging is a fundamental process in development and is executed frequently by developers, reducing the cost of debugging is an essential new challenge. 


To remedy this challenge, APR can use \textit{Small Language Models (SLMs)} and \textit{quantization techniques}.
SLM is also a language model, but it has significantly fewer parameters than LLM, reducing computational and memory requirements for inference and training. 
Quantization converts model weights to lower-bit representations (\eg, converting from float32 to int8), reducing model size and memory usage. While this conversion typically results in some accuracy loss, quantized models can achieve faster inference and lower latency. 
If SLMs and quantization techniques can effectively reduce computational costs while keeping repair accuracy loss to negligible levels, they will enable the practical deployment of LLM-based APR in standard software development environments.



In this paper, we explore 14 SLMs and 2 LLMs, both widely used in APR tasks~\cite{zhang2024usedLLM,Wang2024comprehensivesurveySLM,zhang2024surveyLLM}.
Secondly, we investigate quantization techniques with four bit representations (float32, float16, int8, and int4).
Through an experiment with QuixBugs-Python and QuixBugs-Java~\cite{QuixBugs}, we address the following research questions (RQs):
\begin{itemize}[leftmargin=10mm]
    \item [\textbf{RQ1:}] \rqone{}
    \item [\textbf{RQ2:}] \rqtwo{}
\end{itemize}

Our experiments showed that the best-performing SLMs, such as Phi-3 (3.8B), successfully fixed up to 38 out of 40 bugs, comparable to the performance of the best-performing LLM, Codex, which fixed 39 out of 40 bugs.
We also found that int8 quantization had a minimal impact on repair accuracy, with differences of approximately +0.5 bugs compared to float32, the maximum bit representation.


Our main contributions are as follows:
(1) This is the first comprehensive evaluation of 14 SLMs in APR tasks. We compare the performance of SLMs with LLMs and provide practical insights into the capabilities and limitations of SLMs.
(2) We evaluate the impact of quantization on APR performance, demonstrating that int8 quantization offers a practical trade-off between efficiency and accuracy.
(3) We provide a practical recommendation: code-specific SLMs with int8 quantization can achieve performance comparable to that of LLM-based APR methods, while requiring fewer computational resources.
(4) We provide a comprehensive replication package, including all experimental data and code, to facilitate further research.\footnote{\url{https://doi.org/10.5281/zenodo.15472061}}

\section{Background} \label{background}

This section provides an overview of APR methods, LLM-based APR methods as well as their challenges. Additionally, we introduce SLMs and quantization techniques as potential alternatives to LLMs.

\subsection{APR}
APR methods aim to automate debugging processes. APR methods first identify the buggy location and fix the source code to pass the test cases. 
There exist various APR methods, including heuristic-based~\cite{Le2016SANER}, constraint-based~\cite{Mechtaev2016ICSE}, tempalate-based~\cite{Liu2019ISSTA}, and learning-based approaches~\cite{Zhu2022ESEC/FSE,Jiang2021ICSE}. 
For example, learning-based approaches are based on machine learning techniques such as neural machine translation (NMT)~\cite{Zhu2022ESEC/FSE,Jiang2021ICSE}. These methods are trained on bug-fix examples. Once trained, they can provide modifications to the buggy code.

With the rapid development of machine learning technologies, learning-based APR methods have become the most popular approach. Specifically, LLM-based APR methods attract attention due to their high accuracy and flexibility.
OpenAI Codex~\cite{codex}, for instance, has demonstrated substantial improvements over state-of-the-art APR techniques~\cite{Xia2023ICSE,Prenner2022ICSE}.
While AlphaRepair, the most accurate traditional APR method, achieved around 26\% accuracy, Codex significantly outperformed it with approximately 38\% accuracy~\cite{Xia2023ICSE}.
This highlights the effectiveness of LLM-based approaches in APR tasks, as they can generate high-quality patches for a diverse range of buggy code without requiring additional fine-tuning.

\subsection{Efficiency Challenges of LLM-Based APR}
LLM-based APR methods achieve high accuracy; however, they face non-functional challenges~\cite{Yang2024robustness,Shi2024TOSEM}.
Yang et al.~\cite{Yang2024robustness} investigated 146 papers on code-related tasks using LLMs and identified six important non-functional properties: robustness, security, privacy, explainability, efficiency, and usability.  

In this study, we focus on the challenges of efficiency since they lead to difficulties in applying LLM-based APR methods to a practical software development workflow.
For example, as an efficiency challenge, Yang et al.~\cite{Yang2024robustness} stated that integrating LLM-based approaches into a standard software development workflow is difficult due to the large size of LLMs.
While a model size of 50 MB is considered feasible for integration into modern software development tools such as an integrated development environment (IDE)~\cite{svyatkovskiy2021MSR}, current LLMs far exceed this threshold.
For example, CodeBERT has 125 million parameters and occupies 560 MB of space.\footnote{This value is estimated using the Hugging Face memory estimator~\cite{model-estimator}.}
Such large models lead to high memory consumption and long inference latency, making it difficult for individual developers to use these models in their own environment. 

\subsection{Small Language Models (SLMs)}
To address the efficiency challenges of LLMs, reducing the model size is a promising approach. SLMs therefore emerged as a potential alternative to LLMs. SLMs comprise significantly fewer parameters than LLMs, making them applicable to practical software development workflows. 
For instance, Phi-3 (3.8B parameters)~\cite{Phi-3}, an SLM developed by Microsoft, can be run on a laptop with 16GB of RAM. 
This enables developers to perform APR tasks locally without relying on external servers or preparing large amounts of computational resources.

A possible drawback of SLMs is the performance gap compared to LLMs. Generally, larger models achieve better performance~\cite{Xia2023ICSE}. Hence, in this study, we investigate the trade-offs between model size and performance.


\subsection{Quantization}
Quantization is a technique for compressing models, aiming to reduce computational costs and memory consumption. This approach converts model weights from higher-bit representations (e.g., float32) to lower-bit representations (e.g., int4), thereby reducing the model size. Theoretically, converting from float32 to int8 reduces the model size to approximately one-fourth of its original size.
Moreover, quantization improves inference time, as most modern hardware (CPUs and GPUs) can execute integer operations more quickly than floating point operations.

Since quantization reduces the precision of model weights, it may lead to a loss of information and a decrease in accuracy. Hence, in this paper, we investigate how quantization affects the performance of APR tasks. 
\section{Related Work} \label{RelatedWork}
Various APR methods have been proposed, including template-based~\cite{Zhang2023ASE}, learning-based~\cite{Zhu2022ESEC/FSE}, and LLM-based~\cite{Xia2023ICSE} approaches. Among these, LLM-based APR methods demonstrated approximately 20\% higher accuracy compared to traditional methods~\cite{Xia2023ICSE}, making them a popular area of research in recent years.

However, LLM-based APR methods face several challenges, particularly with regard to high resource consumption~\cite{Yang2024robustness}. For example, Llama2~\cite{Llama2}, one of the largest publicly available models, with up to 70 billion parameters, requires 140 GB of memory and is best run on an eight-GPU configuration (MP=8). To mitigate the high resource consumption, SLMs and quantization techniques, which reduce resource usage, have gained attention.

While SLMs have been explored for code-related tasks, their performance has generally lagged behind that of LLMs~\cite{Lu2024SLMsurvey, Xia2023ICSE}. Xia et al.~\cite{Xia2023ICSE} showed that LLM-based methods significantly outperform SLMs in APR tasks. Since then, several new SLMs, such as Phi-3 (3.8B)\cite{Phi-3}, have been introduced; however, their performance in APR tasks remains underexplored.
Wei et al.~\cite{Wei2023FSE} demonstrated that quantization can reduce resource consumption in code generation tasks without significantly compromising accuracy. However, its impact on APR tasks has not yet been fully examined.
Despite the promising potential of SLMs and quantization, empirical evidence regarding their effectiveness in APR tasks is still lacking. This study, therefore, provides a new empirical evaluation of these methods.

\section{Experimental Design}\label{experiment}

\begin{figure*}[t]
    \centering
    \includegraphics[width=0.9\linewidth]{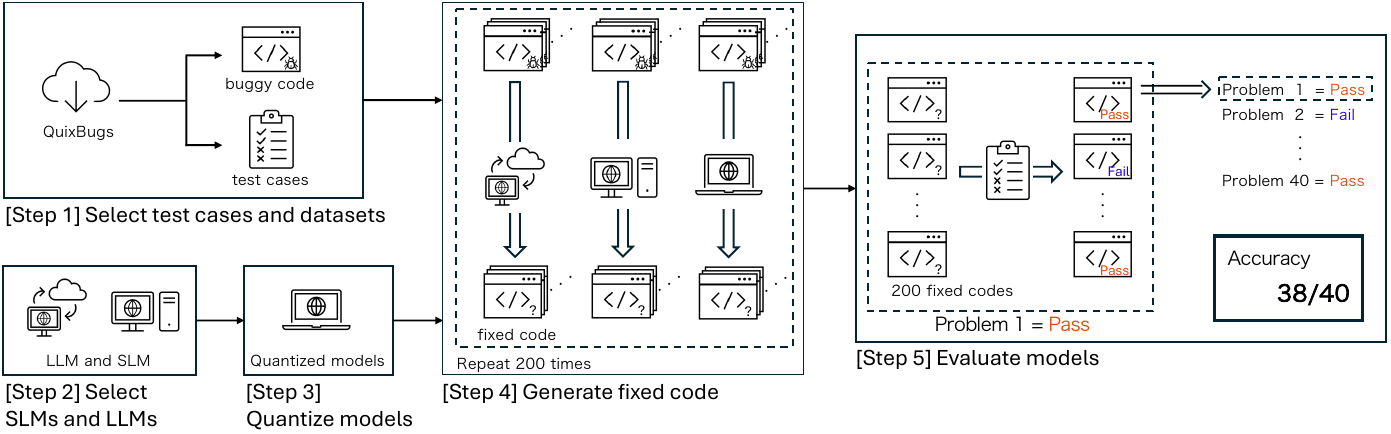}
    \caption{Overall view of the experiment}
    \label{flow}
\end{figure*}

Figure~\ref{flow} illustrates an overview of the steps we follow for our evaluation.
Each step is described in detail below.

\subsection{Select test cases and datasets}
To evaluate the code generation performance, we use the QuixBugs benchmark~\cite{QuixBugs}.
The QuixBugs benchmark~\cite{QuixBugs} is constructed from a programming challenge called Quixey Challenge. In this challenge, programmers were given a program with a single one-line bug and were asked to fix it within one minute. Hence, this dataset targets fixing simple bugs. 

The QuixBugs benchmark comprises 40 buggy methods in Python (\textit{QuixBugs-Python}) and Java (\textit{QuixBugs-Java}). 
Java methods were manually translated from Python methods. 
Hence, this dataset allows evaluation across multiple programming languages.
Listing~\ref{program_python} shows an example of a bug fix in QuixBugs-Python. 
The red line represents the buggy code, while the green line indicates the fixed version. 
This program is a quick sort implementation; the buggy code forgets to include the pivot value in either the less or greater set when splitting a target vector. As a result, the pivot value is lost and not included in the final sorted vector. The fixed code includes the pivot value in the greater set. 

Each buggy method is accompanied by test cases and expected outputs. We use these to evaluate the generated code.

\begin{lstlisting}[language=diff, caption=QuixBugs-Example of Python bug fixes (quick sort), label=program_python]
    def quicksort(arr):
        if not arr:
            return []
               
        pivot = arr[0]
-   greater = quicksort([x for x in arr[1:] if x > pivot])
+   greater = quicksort([x for x in arr[1:] if x >= pivot])
        return lesser + [pivot] + greater
\end{lstlisting}


%

\subsection{Select SLMs and LLMs}
\label{subsec:model}
To generalize the evaluation results and cover a wide range of model sizes, we first selected a variety of language models that cover a wide range of parameter sizes, from small to large.
Based on prior studies~\cite{zhang2024usedLLM,Wang2024comprehensivesurveySLM,zhang2024surveyLLM}, we selected 23 models from 13 different architectures (Table~\ref{model}). 
We refer to an architecture with a specific parameter size as a model. 
To keep the study manageable and aligned with our research goals, we used only a subset of these models in each RQ.
RQ1 aims to evaluate all architectures, while RQ2 focuses on the top-performing models identified in RQ1, specifically the Phi-3 and Qwen2.5-Coder models, to assess the impact of quantization.
The rightmost columns in Table~\ref{model} indicate which models are used in RQ1 and RQ2.

The selected models are described below.

\begin{table}[t]
    \centering
    \caption{Language Model}
    \label{model}
    \scalebox{0.7}{
        \begin{tabular}{lcccccc}
            \toprule
            \textbf{Architecture}                          & \textbf{\# Parameters}         & \textbf{Release}         & \textbf{LLM/SLM}      & Closed & RQ1 & RQ2 \\ \midrule
            GPT-Neo                                 & 125M/1.3B/2.7B                 & 2021/03                  & SLM                  &        & \checkmark    &     \\
            GPT-J                                   & 6B                             & 2021/06                  & SLM                  &        & \checkmark    &     \\
            Phi-2                                   & 2.7B                           & 2023/12                  & SLM                  &        & \checkmark    &     \\
            \multirow{3}{*}{Phi-3}                  & 3.8B                           & 2024/04 & SLM                  &        & \checkmark    &  \checkmark \\
                                                    & \multirow{2}{*}{7B/14B}        & \multirow{2}{*}{2024/04}  & SLM:7B                  &        &     &  \multirow{2}{*}{\checkmark} \\
                                                    &                             &                   & LLM:14B                  &        &     &   \\
            Phi-4-mini-instruct                     & 3.8B                           & 2025/02                  & SLM                  &        & \checkmark    &     \\
            Llama2                                  & 7B                             & 2023/07                  & SLM                  &        & \checkmark    &     \\
            Code Llama                              & 7B                             & 2023/08                  & SLM                  &        & \checkmark    &     \\
            Vicuna                                  & 7B                             & 2023/08                  & SLM                  &        & \checkmark    &     \\
            Llama3                                  & 8B                             & 2024/04                  & SLM                  &        & \checkmark    &     \\
            Llama3.2                                & 1B/3B                          & 2024/09                  & SLM                  &        & \checkmark    &     \\
            \multirow{4}{*}{Qwen2.5-Coder-Instruct} & 1.5B/7B                             & 2024/10                  & SLM                  &        &   &  \checkmark \\
                                                    & 3B                        & 2024/11                  & SLM                  &        &  \checkmark   &  \checkmark \\
                                                    & \multirow{2}{*}{0.5B/14B/32B}  & \multirow{2}{*}{2024/11} & SLM:0.5B                    &        &     &  \multirow{2}{*}{\checkmark} \\
                                                    &                                  &                        & LLM:14B, 32B         &        &     &   \\
            GPT-NeoX                                & 20B                            & 2022/02                  & LLM                  &        & \checkmark    &     \\
            Codex                                   & 12B                            & 2021/10                  & LLM                  &  \checkmark &  \checkmark   &     \\
            \bottomrule
        \end{tabular}
    }
\end{table}

\begin{itemize}
    \item \textbf{GPT-Neo} \cite{gpt-neo}, \textbf{GPT-J} \cite{gpt-j}, \textbf{GPT-NeoX} \cite{gpt-neox}: 
    These three open-source models are based on the GPT-3 Transformer architecture. In this experiment, we used GPT-Neo with parameter sizes of 125M, 1.3B, and 2.7B. GPT-J and GPT-NeoX are larger models with 6B and 20B parameters, respectively. All of these models are trained on The Pile~\cite{Pile}, an 800GB dataset consisting of 22 diverse text-based datasets, which includes 7.6\% of GitHub code.
    \item \textbf{Codex}\cite{codex}: 
    A 12B-parameter GPT-3-based model optimized for code generation. Codex is initialized with GPT-3 weights trained on a natural language corpus and subsequently fine-tuned on a 159GB corpus of code files.
    \item \textbf{Llama2} \cite{Llama2}, \textbf{Code Llama} \cite{CodeLlama}, \textbf{Llama3} \cite{Llama3}, \textbf{Llama3.2} \cite{Llama3.2}: 
    Llama2 is a text generation model pre-trained on 2 trillion tokens.
	Code Llama is a specialized variant of Llama2, fine-tuned for code generation tasks. It can generate both code and natural language descriptions of code based on given inputs. The model has 7B parameters.
	Llama3 is a pre-trained model trained on over 15 trillion tokens. Its training dataset is seven times larger than that of Llama2 and contains four times more code data.
	Llama3.2 is a lightweight and efficient model in the Llama3 series, available in 1B and 3B parameter configurations.
    \item \textbf{Vicuna} \cite{Vicuna}: 
    A chat assistant model fine-tuned from Llama2 using user-shared conversations collected from ShareGPT~\cite{ShareGPT}.
    \item \textbf{Phi-2} \cite{Phi-2}, \textbf{Phi-3} \cite{Phi-3}, \textbf{Phi-4-mini-instruct} \cite{Phi-4-mini-instruct}: 
    Transformer-based language models developed by Microsoft. Phi-3 and Phi-4-mini-instruct are successors to Phi-2.
    \item \textbf{Qwen-2.5-coder} \cite{Qwen2.5-Coder}: 
    A code-specific variant of Qwen-2.5, a language model developed by the Qwen team at Alibaba Cloud.
	In RQ1, a 3B-parameter model was used.
	In RQ2, six different models with parameter sizes of 0.5B, 1.5B, 3B, 7B, 14B, and 32B were employed.
\end{itemize}


Since there is no clear threshold to distinguish between SLMs and LLMs, and the definitions vary across studies~\cite{Lu2024SLMsurvey,Wang2024comprehensivesurveySLM}, we define SLMs as models capable of running inference on a home GPU.
Specifically, we consider NVIDIA GeForce RTX 30 series GPUs, which hold a significant market share among home GPUs.
The VRAM capacities of these GPUs are 6GB, 8GB, 12GB, and 24GB.
Therefore, we define SLMs as models that require less than 24GB of VRAM for inference.
For example, a model with approximately 10B parameters with float16 bit representation can be run inference on a GPU with 24GB of VRAM.

We used models available on Hugging Face to load weights and generate outputs locally. 
In contrast, Codex was accessed via the OpenAI API, as it is not publicly available.

The parameters for these models were set as follows
\begin{itemize}
    \item \textbf{top-p=0.95}:
    Limits the candidate tokens considered during selection.
    \item \textbf{temperature=0.8}:
    Adds randomness to the token selection.
    \item \textbf{number of samples per bug=200}: 
    Specifies the number of candidate patches generated for each bug.
\end{itemize}


These settings are consistent with those used in previous studies on LLM-based APR~\cite{Xia2023ICSE,codex}.


\subsection{Quantize models}
There exist various quantization methods. In this study, we employ the \textit{GPTQ quantization technique}~\cite{frantar2023gptqaccurateposttrainingquantization}. 
It has been widely adopted in previous studies~\cite{zhao2025arxiv,zheng2024arxiv} and is easily accessible through the Hugging Face Transformers library.
With this method, we evaluate the following four model weights in our experiments:
\begin{itemize}
    \item \textbf{float32}: 
    Single-precision floating-point format. This is the default precision setting, offering the highest accuracy but requiring significant memory and computational resources.
    \item \textbf{float16}: 
    Half-precision floating-point format. This format reduces memory usage by approximately 50\% compared to float32.
    \item \textbf{int8}: 
    8-bit fixed-point format. This format provides even lower memory consumption than float16, though it may lead to inference accuracy loss. 
    \item \textbf{int4}: 
    Highly compressed 4-bit precision format. This format is significantly memory-efficient and suitable for low-resource environments. However, it may introduce significant accuracy loss.
\end{itemize}


Table~\ref{required_memory_example} presents an example of the estimated memory requirements for the Qwen-2.5-coder-7B-Instruct model under various quantization formats. Each cell displays the required memory as estimated by the Hugging Face model memory estimator~\cite{model-estimator}.
To illustrate the relationship between these memory requirements and NVIDIA GeForce RTX 30 series GPUs, we have highlighted cells in green (requiring at least 6GB), orange (at least 8GB), and red (at least 24GB). We do not highlight cells corresponding to 12GB since a 12GB GPU supports the same quantization formats as an 8GB GPU in this case. 

Table~\ref{required_memory_example} shows that memory requirements vary significantly across different quantization formats, leading to distinct GPU specifications. 
For instance, the float16 format requires 24GB of memory, whereas the int4 format requires only 6GB. 


In this study, we simulate four different environments in which developers use GPUs with 24GB, 12GB, 8GB, or 6GB of memory.
We apply quantization techniques to a model, resulting in four different quantized versions: float32, float16, int8, and int4.
For each GPU memory environment, we select the largest quantized model that can still be executed on that GPU.

For example, when two quantized models are available with sizes of 23GB and 20GB, respectively, both can be executed on a 24GB GPU. In this case, the 23GB model is selected, as it utilizes the maximum available GPU memory. Following this policy, we prepared four distinct quantized SLMs, each corresponding to a specific GPU memory environment.

\begin{table}[t]
    \centering
    \caption{Estimated required memory for each quantization setting [GB]}
    \label{required_memory_example}
    \scalebox{0.9}{
    \begin{tabular}{lcccc}
    \toprule
    \textbf{Model}              & \textbf{float32}             & \textbf{float16}             & \textbf{int8}                & \textbf{int4}                \\ \midrule
    Qwen2.5-Coder-7B-Instruct   & 27.22                         & \cellcolor[HTML]{FF9999}13.61 & \cellcolor[HTML]{FFCC67}6.8   & \cellcolor[HTML]{32CB00}3.4                           \\ \bottomrule
    \end{tabular}
    }
\end{table}

\subsection{Generate fixed code}
We use the selected LLMs, SLMs, and quantized SLMs to generate fixed code for the buggy programs in QuixBugs. 
In this study, we treat our APR task as a \textit{complete function generation task}~\cite{Xia2023ICSE}. 
Under this setting, the input is a buggy method, and the goal is to generate a fixed version of that method, rather than generating a method from scratch.

Since our selected language models are not pre-trained for APR, providing the buggy function alone does not yield a fixed version. 
To enable these models to perform APR effectively, we employ a task-specific prompt designed to facilitate few-shot learning, following the prompt structure used in a previous study~\cite{Xia2023ICSE}. 
This prompt primarily includes two bug-fix examples: the first is a conceptual example demonstrating the task (using a function that computes Fibonacci numbers), and the second is a real example from the QuixBugs dataset that helps the model learn the coding style used in QuixBugs.

\begin{figure}[t]
    \centering
    \includegraphics[scale=0.5]{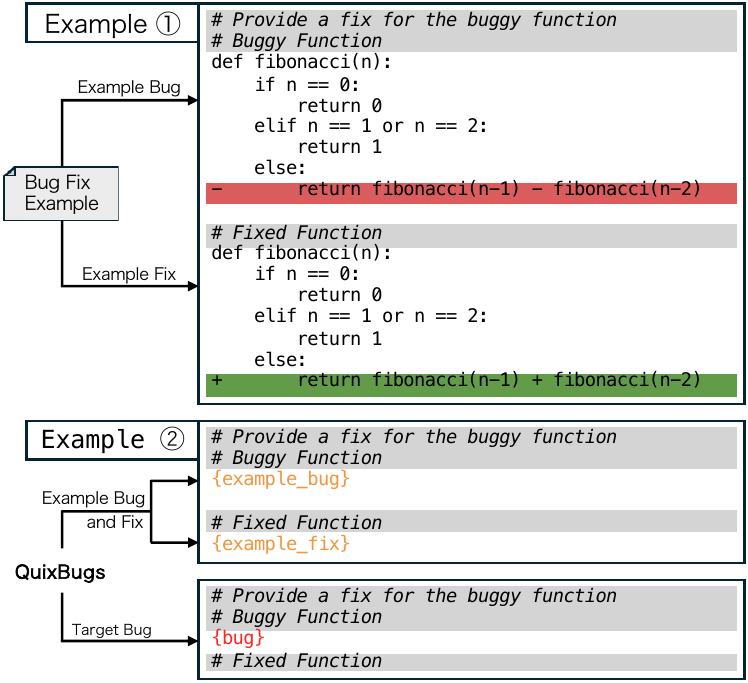}
    \caption{Prompt for complete function generation}
    \label{prompt}
\end{figure}

Figure~\ref{prompt} illustrates the prompt template used in this study. 
The prompt begins with a natural language description of the APR task, clearly explaining what is expected from the model. 
This is followed by the two examples described above, where each example includes a buggy function and its corresponding fixed version, with each function prefixed by a comment indicating whether it is buggy or fixed. 
Finally, the prompt provides a function that is to be repaired.

Because language models generate outputs randomly, the same input may yield different results on different runs. 
To account for this randomness, we repeated the generation process 200 times for each buggy method, resulting in 200 fixed code samples for each instance.




\subsection{Evaluate models}
We evaluate the generated code using the test cases provided in the QuixBugs dataset. A bug is considered fixed if at least one out of the 200 generated code samples passes all test cases; otherwise, it is marked as a failure. Since the QuixBugs dataset contains 40 bugs, we report the number of bugs successfully fixed by each model. For example, if Phi-3 fixes 38 out of 40 bugs, its performance is expressed as 38/40.



\section{Results} \label{result}

\subsection{RQ1: \rqone{}}

In RQ1, we provided a comparative analysis of performance between different types of LLMs and SLMs in APR. 
Specifically, we investigated the effectiveness of 2 LLMs and 14 SLMs on two benchmarks.

\textbf{Most SLMs exhibit high accuracy, performing comparably or even surpassing LLMs.}
Table~\ref{RQ1result} shows the number of successfully fixed bugs (out of 40) for each model, across both the QuixBugs-Python and QuixBugs-Java datasets, with the best-performing SLMs and LLMs highlighted in bold. The table also includes the parameter size, release date, and \textit{code related}.
Here, \textit{code-related} refers to models that were trained or fine-tuned on code datasets, such as those containing programming tasks. 
For example, Code Llama is categorized as \textit{code-related} models due to its training objectives specifically tailored for code understanding and generation.

For QuixBugs-Python, Codex (LLM) achieved the highest accuracy by fixing 39 out of 40 bugs. Among SLMs, both Phi-3 (3.8B) and Qwen2.5-Coder-3B-Instruct demonstrated comparable performance, fixing 38 bugs each, nearly matching the accuracy of Codex. 
Notably, 9 SLMs fixed more than 20 bugs, outperforming GPT-NeoX (20B), the largest LLM in RQ1, which fixed only 20 out of 40 bugs.
A similar trend appears in QuixBugs-Java.


\textbf{There is a correlation between model size and bug fixes in the same model series.} 
A positive correlation exists between model size and bug-fixing accuracy within the same model series. 
As seen in Table~\ref{RQ1result}, larger models in the same family (e.g., GPT-Neo and Llama3.2) show better accuracy. 
For instance, in the GPT-Neo series, the number of fixed bugs in QuixBugs-Python increases steadily with model size, from 2 (125M) to 3 (1.3B) to 6 (2.7B). 
This finding aligns with previous research in APR~\cite{Xia2023ICSE}.
\textbf{Models designed and fine-tuned for code-related tasks demonstrate superior bug-fixing performance.}
Codex, Code Llama, and Qwen2.5-Coder demonstrated this by fixing 39, 37, and 38 bugs respectively in QuixBugs-Python, significantly outperforming general-purpose models. 
A clear example is Code Llama, a code-specialized version of Llama2, which improved its bug-fixing accuracy from 23 bugs (Llama2) to 37 bugs (Code Llama).
In contrast, Vicuna, a version of Llama2 fine-tuned for conversation, fixed 19 fewer bugs than Llama2. 
These suggest that while fine-tuning models for code-related tasks improves their bug-fixing abilities, fine-tuning for general conversation actually reduces their program repair capabilities.

\begin{tcolorbox}\textbf{RQ1 Summary:}
Nine of the 14 SLMs outperformed the largest LLM (GPT-NeoX), with the best-performing SLMs (Phi-3 and Qwen2.5-Coder-3B-Instruct) fixing 38 out of 40 bugs, a performance comparable to that of Codex, the best LLM.
Moreover, code-specialized models demonstrated superior bug-fixing performance compared to general-purpose models, such as Code Llama versus Llama2.
These findings highlight the effectiveness of SLMs in APR, particularly when designed and fine-tuned for code-related tasks.

\end{tcolorbox}



\begin{table}[t]
    \centering
    \caption{
    Number of fixed bugs in QuixBugs
    }
    \label{RQ1result}
    \scalebox{0.75}{
        \begin{tabular}{lccccc}
            \toprule
            \textbf{Model}   & \textbf{\#Parameters}  & \textbf{Release}                  & \makecell{\textbf{QuixBugs}\\\textbf{Python}} & \makecell{\textbf{QuixBugs}\\\textbf{Java}} & \makecell{\textbf{Code}\\\textbf{related}}\\ \midrule
            GPT-Neo & 125M & 2021/03                                                      & 2                        & 1                      \\
            GPT-Neo & 1.3B & 2021/03                                                       & 3                        & 4                      \\
            GPT-Neo & 2.7B & 2021/03                                                       & 6                        & 7                      \\
            GPT-J & 6B & 2021/06                                                             & 14                       & 6                      \\
            Phi-2 & 2.7B & 2023/12                                                              & 33                       & 18                     \\
            Phi-3 & 3.8B & 2024/04                                                             & \uline{\textbf{38}}     & 29                     \\
            Phi-4-mini-instruct & 3.8B & 2025/02                                                        & 37                       &  \uline{\textbf{30}}                    \\
            Llama2 & 7B & 2023/07                                                             & 23                       & 6                      \\
            Code Llama & 7B & 2023/08                                                         & 37                       & \uline{\textbf{30}}   & \checkmark                  \\
            Vicuna & 7B & 2023/08                                                            & 19                       & 6                      \\
            Llama3 & 8B & 2024/04                                                            & 36                       & 25                     \\
            Llama3.2 & 1B & 2024/09                                                        & 28                       & 17                     \\
            Llama3.2 & 3B & 2024/10                                                        & 35                       & 23                     \\ 
            \begin{tabular}[c]{@{}l@{}}Qwen2.5-Coder\\ -3B-Instruct\end{tabular}& 3B & 2024/11 & \uline{\textbf{38}}     & 29   & \checkmark                  \\ \midrule
            GPT-NeoX & 20B & 2022/02                                                           & 20                       & 6                      \\
            Codex & 12B & 2021/10                                                             & \uline{\textbf{39}}      & \uline{\textbf{30}}    & \checkmark                 \\ \bottomrule
            \end{tabular}
    }
\end{table}

\subsection{RQ2: \rqtwo{}}


In RQ2, we investigated how quantization affects the bug-fixing accuracy of the Phi-3 and Qwen2.5-Coder series, the best-performing SLMs in RQ1. 
Table~\ref{required_memory} shows the memory requirements for the four quantization settings (float32, float16, int8, and int4) during model inference.
Table~\ref{required_memory_example}, \ref{required_memory} and~\ref{RQ2result} use color coding to indicate the memory constraints of home GPUs: green ($\leq$ 6GB), orange ($\leq$ 8GB), and red ($\leq$ 24GB).
We do not highlight cells corresponding to 12GB, as a 12GB GPU supports the same quantization formats as either 8GB or 24GB GPU.


\textbf{Quantization up to int8 not only preserves accuracy but also significantly reduces memory usage.} 
Table~\ref{RQ2result} shows the number of fixed bugs under four different quantization settings. 
The numbers in parentheses indicate the change in fixed bugs compared to the full-precision setting (float32). 
Using float16 quantization resulted in fixing 0.5 more bugs on average compared to float32. 
With int8 quantization, the models fixed 0.25 fewer bugs on average. 
However, int4 quantization led to a significant decrease, fixing 11.25 fewer bugs on average compared to float32. 
While quantization up to int8 minimally affects accuracy, int4 substantially reduces it.

\begin{tcolorbox}\textbf{RQ2 Summary:}
Compared to full precision (float32), the bug-fixing performance of SLMs showed minimal changes: a slight increase of 0.5 bugs for float16 and a small decrease of 0.25 bugs for int8. However, int4 quantization significantly reduced performance, with an average decrease of 11.25 bugs.
\end{tcolorbox}

\begin{table}[tbp]
  \centering
    \caption{Estimated required memory for each quantization setting [GB]}
    \label{required_memory}
    \scalebox{0.8}{
    \begin{tabular}{lcccc}
    \toprule
    \textbf{Model}              & \textbf{float32}             & \textbf{float16}             & \textbf{int8}                & \textbf{int4}                \\ \midrule
    Phi-3 3.8B                  & \cellcolor[HTML]{FF9999}14.23 & \cellcolor[HTML]{FFCC67}7.12  & \cellcolor[HTML]{32CB00}3.56  & 1.78                          \\
    Phi-3 7B                    & 26                            & \cellcolor[HTML]{FF9999}13    & \cellcolor[HTML]{FFCC67}6.5   & \cellcolor[HTML]{32CB00}3.25  \\
    Phi-3 14B                   & 52.01                         & 26                            & \cellcolor[HTML]{FF9999}13    & \cellcolor[HTML]{FFCC67}6.5   \\
    Qwen2.5-Coder-0.5B-Instruct & \cellcolor[HTML]{32CB00}2.22  & 1.11                          & 0.57                          & 0.28                          \\
    Qwen2.5-Coder-1.5B-Instruct & \cellcolor[HTML]{FFCC67}6.63  & \cellcolor[HTML]{32CB00}3.31  & 1.66                          & 0.85                          \\
    Qwen2.5-Coder-3B-Instruct   & \cellcolor[HTML]{FF9999}12.62 & \cellcolor[HTML]{FFCC67}6.31  & \cellcolor[HTML]{32CB00}3.16  & 1.58                          \\
    Qwen2.5-Coder-7B-Instruct   & 27.22                         & \cellcolor[HTML]{FF9999}13.61 & \cellcolor[HTML]{FFCC67}6.8   & \cellcolor[HTML]{32CB00}3.4   \\
    Qwen2.5-Coder-14B-Instruct  & 53.62                         & 26.81                         & \cellcolor[HTML]{FF9999}13.41 & \cellcolor[HTML]{FFCC67}6.7   \\
    Qwen2.5-Coder-32B-Instruct  & 121.15                        & 60.58                         & 30.29                         & \cellcolor[HTML]{FF9999}15.14 \\ \bottomrule
    \end{tabular}
    }
    \vspace{2mm}\\
\textbf{Minimum required memory:}
\scalebox{0.8}{
\begin{tabular}{lll}
\cellcolor[HTML]{32CB00}~$\leq$6GB &
\cellcolor[HTML]{FFCC67}~$\leq$8GB &
\cellcolor[HTML]{FF9999}~$\leq$24GB
\end{tabular}}
  \vspace{1em} 
    \caption{Accuracy for each quantization setting}
    \label{RQ2result}
    \scalebox{0.8}{
    \begin{tabular}{lcccc}
    \toprule
    \textbf{Model}              & \textbf{float32}          & \textbf{float16}                                                           & \textbf{int8}                       & \textbf{int4}              \\ \midrule
    Phi-3 3.8B                  & \cellcolor[HTML]{FF9999}35 & \cellcolor[HTML]{FFCC67}{\color[HTML]{333333} 38( {\color{red}+3})}       & \cellcolor[HTML]{32CB00}{\color[HTML]{333333} 37( {\color{red}+2})}                 & 36({\color{red}+1})                     \\
    Phi-3 7B                    &                            & \cellcolor[HTML]{FF9999}39                                                  & \cellcolor[HTML]{FFCC67}39           & \cellcolor[HTML]{32CB00}39                         \\
    Phi-3 14B                   &                            &                                                                             & \cellcolor[HTML]{FF9999}39           & \cellcolor[HTML]{FFCC67}37 \\
    Qwen2.5-Coder-0.5B-Instruct & \cellcolor[HTML]{32CB00}27 & 30({\color{red}+3})                                                        & 28({\color{red}+1})                        & 25({\color{blue}-2})                     \\
    Qwen2.5-Coder-1.5B-Instruct & \cellcolor[HTML]{FFCC67}38 & \cellcolor[HTML]{32CB00}{\color[HTML]{333333} 34( {\color{blue}-4})}                                                                     & 34({\color{blue}-4})                             & 32({\color{blue}-6})                     \\
    Qwen2.5-Coder-3B-Instruct   & \cellcolor[HTML]{FF9999}38 & \cellcolor[HTML]{FFCC67}38(±0)                                            & \cellcolor[HTML]{32CB00}38(±0)                              & 0({\color{blue}-38})                     \\
    Qwen2.5-Coder-7B-Instruct   &                            & \cellcolor[HTML]{FF9999}38                                                  & \cellcolor[HTML]{FFCC67}39          & \cellcolor[HTML]{32CB00}38                         \\
    Qwen2.5-Coder-14B-Instruct  &                            &                                                                             & \cellcolor[HTML]{FF9999}38         & \cellcolor[HTML]{FFCC67}38 \\
    Qwen2.5-Coder-32B-Instruct  &                            &                                                                              &                                     & \cellcolor[HTML]{FF9999}35 \\
    \midrule
    Ave.diff                    & --                         & {\color{red}+0.5}           & {\color{blue}-0.25}           & {\color{blue}-11.25}           \\
    \bottomrule
    \end{tabular}
    }
  \label{tab:rq2_combined}
\end{table}



\section{Threats to Validity} \label{sec:threats}

\noindent \textbf{Internal Validity:}
In this study, we evaluated the performance of LLMs and SLMs using the QuixBugs dataset. 
If the training data of these models includes the QuixBugs data, the models may generate correct patches by simply reproducing the training data. 
This data leakage could lead to an overestimation of their performance. 
Future studies should incorporate additional datasets to avoid this threat to validity.


\smallskip \noindent \textbf{External Validity:}
We evaluated 14 SLMs and 2 LLMs on a repair dataset across two programming languages and examined the impact of quantization on 9 SLMs. 
While this is a large-scale empirical study providing insights into the performance of SLMs and LLMs in APR, our findings may not generalize to other models or programming languages. 
Future studies should evaluate the performance of additional LLMs and SLMs across other programming languages.

Our evaluation is limited to the QuixBugs dataset. 
This limitation affects the generalizability of our findings. 



\section{Conclusion} \label{conclusion}
This study evaluated the impact of SLMs on the APR task. While Codex, an LLM, fixed 39 of 40 bugs, the best-performing SLMs, Phi-3 and Qwen2.5-Coder-3B-Instruct, fixed 38 of 40 bugs, which is comparable to the best-performing LLM. Moreover, float16 and int8 quantization had minimal impact on SLM performance, with differences of only +0.5 and –0.25 bugs fixed, respectively, compared to float32. These small performance differences, combined with the reduced computational demands, make SLMs feasible for practical software development environments, such as those using NVIDIA GeForce RTX 30 series GPUs.
Our practical recommendation is that \textbf{code-specific SLMs with int8 quantization can achieve performance comparable to LLM-based APR methods while requiring significantly less computational resources}.


For future work, we aim to assess the generalizability and robustness of SLM-based APR methods. We plan to evaluate these models on other datasets, such as SWE-Bench~\cite{jimenez2024ICLR}, to consider other real-world scenarios. Also, enhancing SLM performance remains a promising direction, for instance by incorporating project-specific context through fine-tuning.

\section*{Acknowledgment}
We gratefully acknowledge the financial support of: (1) JSPS for the KAKENHI grants (JP24K02921, JP25K03100); (2) Japan Science and Technology Agency (JST) as part of Adopting Sustainable Partnerships for Innovative Research Ecosystem (ASPIRE), Grant NumberJPMJAP2415, and (3) the Inamori Research Institute for Science for supporting Yasutaka Kamei via the InaRIS Fellowship.

\bibliographystyle{IEEEtran}
\bibliography{references_short}

\begin{thebibliography}{10}
\providecommand{\url}[1]{#1}
\csname url@samestyle\endcsname
\providecommand{\newblock}{\relax}
\providecommand{\bibinfo}[2]{#2}
\providecommand{\BIBentrySTDinterwordspacing}{\spaceskip=0pt\relax}
\providecommand{\BIBentryALTinterwordstretchfactor}{4}
\providecommand{\BIBentryALTinterwordspacing}{\spaceskip=\fontdimen2\font plus
\BIBentryALTinterwordstretchfactor\fontdimen3\font minus
  \fontdimen4\font\relax}
\providecommand{\BIBforeignlanguage}[2]{{%
\expandafter\ifx\csname l@#1\endcsname\relax
\typeout{** WARNING: IEEEtran.bst: No hyphenation pattern has been}%
\typeout{** loaded for the language `#1'. Using the pattern for}%
\typeout{** the default language instead.}%
\else
\language=\csname l@#1\endcsname
\fi
#2}}
\providecommand{\BIBdecl}{\relax}
\BIBdecl

\bibitem{britton2013reversible}
T.~Britton, L.~Jeng, G.~Carver, P.~Cheak, and T.~Katzenellenbogen, ``Reversible
  debugging software-quantify the time and cost saved using reversible
  debuggers,'' \emph{Univ. Cambridge, Cambridge, UK}, 2013.

\bibitem{Shariffeen2021TOSEM}
R.~S. Shariffdeen, S.~H. Tan, M.~Gao, and A.~Roychoudhury, ``Automated patch
  transplantation,'' \emph{ACM Trans. Softw. Eng. Methodol.}, vol.~30, no.~1,
  2021.

\bibitem{Ye2021ESE}
H.~Ye, M.~Martinez, and M.~Monperrus, ``Automated patch assessment for program
  repair at scale,'' \emph{Empir. Softw. Eng.}, vol.~26, no.~20, 2021.

\bibitem{Weimer2009ICSE}
W.~Weimer, T.~Nguyen, C.~Le~Goues, and S.~Forrest, ``Automatically finding
  patches using genetic programming,'' in \emph{Proc. ICSE 2009}, ser. ICSE
  '09.\hskip 1em plus 0.5em minus 0.4em\relax USA: IEEE Computer Society, 2009,
  p. 364–374.

\bibitem{Xia2023ICSE}
C.~S. Xia, Y.~Wei, and L.~Zhang, ``Automated program repair in the era of large
  pre-trained language models,'' in \emph{Proc. ICSE 2023}.\hskip 1em plus
  0.5em minus 0.4em\relax IEEE, 2023, pp. 1482--1494.

\bibitem{Prenner2022ICSE}
J.~A. Prenner, H.~Babii, and R.~Robbes, ``Can openai's codex fix bugs? an
  evaluation on quixbugs,'' in \emph{Proc. ICSEW 2022}, ser. APR '22.\hskip 1em
  plus 0.5em minus 0.4em\relax New York, NY, USA: Association for Computing
  Machinery, 2022, p. 69–75.

\bibitem{Fan2023ICSE}
Z.~Fan, X.~Gao, M.~Mirchev, A.~Roychoudhury, and S.~H. Tan, ``Automated repair
  of programs from large language models,'' in \emph{Proc. ICSE 2023}, ser.
  ICSE '23.\hskip 1em plus 0.5em minus 0.4em\relax IEEE Press, 2023, p.
  1469–1481.

\bibitem{Yang2024robustness}
\BIBentryALTinterwordspacing
Z.~Yang, Z.~Sun, T.~Z. Yue, P.~Devanbu, and D.~Lo, ``Robustness, security,
  privacy, explainability, efficiency, and usability of large language models
  for code,'' 2024. [Online]. Available: \url{https://arxiv.org/abs/2403.07506}
\BIBentrySTDinterwordspacing

\bibitem{Huang2025ICSE}
K.~Huang, J.~Zhang, X.~Meng, and Y.~Liu, ``{ Template-Guided Program Repair in
  the Era of Large Language Models },'' in \emph{Proc. ICSE 2025}.\hskip 1em
  plus 0.5em minus 0.4em\relax Los Alamitos, CA, USA: IEEE Computer Society,
  May 2025, pp. 367--379.

\bibitem{zhang2024usedLLM}
\BIBentryALTinterwordspacing
Q.~Zhang, C.~Fang, Y.~Xie, Y.~Ma, W.~Sun, Y.~Yang, and Z.~Chen, ``A systematic
  literature review on large language models for automated program repair,''
  2024. [Online]. Available: \url{https://arxiv.org/abs/2405.01466}
\BIBentrySTDinterwordspacing

\bibitem{Wang2024comprehensivesurveySLM}
\BIBentryALTinterwordspacing
F.~Wang, Z.~Zhang, X.~Zhang, Z.~Wu, T.~Mo, Q.~Lu, and e.~a. Wanjing~Wang, ``A
  comprehensive survey of small language models in the era of large language
  models: Techniques, enhancements, applications, collaboration with llms, and
  trustworthiness,'' 2024. [Online]. Available:
  \url{https://arxiv.org/abs/2411.03350}
\BIBentrySTDinterwordspacing

\bibitem{zhang2024surveyLLM}
\BIBentryALTinterwordspacing
Q.~Zhang, C.~Fang, Y.~Xie, Y.~Zhang, Y.~Yang, W.~Sun, S.~Yu, and Z.~Chen, ``A
  survey on large language models for software engineering,'' 2024. [Online].
  Available: \url{https://arxiv.org/abs/2312.15223}
\BIBentrySTDinterwordspacing

\bibitem{QuixBugs}
D.~Lin, J.~Koppel, A.~Chen, and A.~Solar-Lezama, ``Quixbugs: a multi-lingual
  program repair benchmark set based on the quixey challenge,'' in \emph{Proc.
  SPLASH 2017}, ser. SPLASH Companion 2017.\hskip 1em plus 0.5em minus
  0.4em\relax New York, NY, USA: Association for Computing Machinery, 2017, p.
  55–56.

\bibitem{Le2016SANER}
X.~B.~D. Le, D.~Lo, and C.~Le~Goues, ``History driven program repair,'' in
  \emph{Proc. SANER 2016}, vol.~1.\hskip 1em plus 0.5em minus 0.4em\relax IEEE,
  2016, pp. 213--224.

\bibitem{Mechtaev2016ICSE}
S.~Mechtaev, J.~Yi, and A.~Roychoudhury, ``Angelix: Scalable multiline program
  patch synthesis via symbolic analysis,'' in \emph{Proc. ICSE 2016}, 2016, pp.
  691--701.

\bibitem{Liu2019ISSTA}
K.~Liu, A.~Koyuncu, D.~Kim, and T.~F. Bissyandé, ``Tbar: revisiting
  template-based automated program repair,'' in \emph{Proc. ISSTA 2019}, ser.
  ISSTA ’19.\hskip 1em plus 0.5em minus 0.4em\relax ACM, Jul. 2019.

\bibitem{Zhu2022ESEC/FSE}
Q.~Zhu, Z.~Sun, Y.-a. Xiao, W.~Zhang, K.~Yuan, Y.~Xiong, and L.~Zhang, ``A
  syntax-guided edit decoder for neural program repair,'' in \emph{Proc.
  ESEC/FSE 2023}, ser. ESEC/FSE 2021.\hskip 1em plus 0.5em minus 0.4em\relax
  New York, NY, USA: Association for Computing Machinery, 2021, p. 341–353.

\bibitem{Jiang2021ICSE}
N.~Jiang, T.~Lutellier, and L.~Tan, ``Cure: Code-aware neural machine
  translation for automatic program repair,'' in \emph{Proc. ICSE 2021}.\hskip
  1em plus 0.5em minus 0.4em\relax IEEE, May 2021, p. 1161–1173.

\bibitem{codex}
\BIBentryALTinterwordspacing
M.~Chen, J.~Tworek, H.~Jun, Q.~Yuan, H.~P. de~Oliveira~Pinto, J.~Kaplan, and
  e.~a. Harri~Edwards, ``Evaluating large language models trained on code,''
  2021. [Online]. Available: \url{https://arxiv.org/abs/2107.03374}
\BIBentrySTDinterwordspacing

\bibitem{Shi2024TOSEM}
J.~Shi, Z.~Yang, and D.~Lo, ``Efficient and green large language models for
  software engineering: Vision and the road ahead,'' \emph{ACM Trans. Softw.
  Eng. Methodol.}, Dec. 2024, just Accepted.

\bibitem{svyatkovskiy2021MSR}
A.~Svyatkovskiy, S.~Lee, A.~Hadjitofi, M.~Riechert, J.~V. Franco, and
  M.~Allamanis, ``Fast and memory-efficient neural code completion,'' in
  \emph{Proc. MSR 2021}.\hskip 1em plus 0.5em minus 0.4em\relax IEEE, 2021, pp.
  329--340.

\bibitem{model-estimator}
H.~F.~M. memory estimator,
  \url{https://huggingface.co/docs/accelerate/usage_guides/model_size_estimator}
  [2024-01-17].

\bibitem{Phi-3}
M.~Abdin, J.~Aneja, H.~Awadalla, A.~Awadallah, A.~A. Awan, N.~Bach, and e.~a.
  Bahree, Amit, ``Phi-3 technical report: A highly capable language model
  locally on your phone,'' \emph{arXiv preprint arXiv:2404.14219}, 2024.

\bibitem{Zhang2023ASE}
Q.~Zhang, C.~Fang, T.~Zhang, B.~Yu, W.~Sun, and Z.~Chen, ``{ Gamma: Revisiting
  Template-Based Automated Program Repair Via Mask Prediction },'' in
  \emph{Proc. ASE 2023}.\hskip 1em plus 0.5em minus 0.4em\relax Los Alamitos,
  CA, USA: IEEE Computer Society, Sep. 2023, pp. 535--547.

\bibitem{Llama2}
\BIBentryALTinterwordspacing
H.~Touvron, L.~Martin, K.~Stone, P.~Albert, A.~Almahairi, Y.~Babaei, and e.~a.
  Nikolay~Bashlykov, ``Llama 2: Open foundation and fine-tuned chat models,''
  2023. [Online]. Available: \url{https://arxiv.org/abs/2307.09288}
\BIBentrySTDinterwordspacing

\bibitem{Lu2024SLMsurvey}
\BIBentryALTinterwordspacing
Z.~Lu, X.~Li, D.~Cai, R.~Yi, F.~Liu, X.~Zhang, and e.~a. Nicholas D.~Lane,
  ``Small language models: Survey, measurements, and insights,'' 2024.
  [Online]. Available: \url{https://arxiv.org/abs/2409.15790}
\BIBentrySTDinterwordspacing

\bibitem{Wei2023FSE}
X.~Wei, S.~K. Gonugondla, S.~Wang, W.~Ahmad, B.~Ray, H.~Qian, and e.~a. Li,
  Xiaopeng, ``Towards greener yet powerful code generation via quantization: An
  empirical study,'' in \emph{Proc. ESEC/FSE 2023}, 2023, pp. 224--236.

\bibitem{gpt-neo}
S.~Black, G.~Leo, P.~Wang, C.~Leahy, and S.~Biderman, ``Gpt-neo: Large scale
  autoregressive language modeling with mesh-tensorflow,'' Aug. 2021.

\bibitem{gpt-j}
B.~Wang and A.~Komatsuzaki, ``{GPT-J-6B: A 6 Billion Parameter Autoregressive
  Language Model},'' \url{https://github.com/kingoflolz/mesh-transformer-jax},
  May 2021.

\bibitem{gpt-neox}
\BIBentryALTinterwordspacing
S.~Black, S.~Biderman, E.~Hallahan, Q.~Anthony, L.~Gao, L.~Golding, and e.~a.
  Horace~He, ``Gpt-neox-20b: An open-source autoregressive language model,''
  2022. [Online]. Available: \url{https://arxiv.org/abs/2204.06745}
\BIBentrySTDinterwordspacing

\bibitem{Pile}
\BIBentryALTinterwordspacing
L.~Gao, S.~Biderman, S.~Black, L.~Golding, T.~Hoppe, C.~Foster, and e.~a.
  Jason~Phang, ``The pile: An 800gb dataset of diverse text for language
  modeling,'' 2020. [Online]. Available: \url{https://arxiv.org/abs/2101.00027}
\BIBentrySTDinterwordspacing

\bibitem{CodeLlama}
\BIBentryALTinterwordspacing
B.~Rozière, J.~Gehring, F.~Gloeckle, S.~Sootla, I.~Gat, X.~E. Tan, and e.~a.
  Yossi~Adi, ``Code llama: Open foundation models for code,'' 2024. [Online].
  Available: \url{https://arxiv.org/abs/2308.12950}
\BIBentrySTDinterwordspacing

\bibitem{Llama3}
\BIBentryALTinterwordspacing
A.~Grattafiori, A.~Dubey, A.~Jauhri, A.~Pandey, A.~Kadian, A.~Al-Dahle, and
  e.~a. Aiesha~Letman, ``The llama 3 herd of models,'' 2024. [Online].
  Available: \url{https://arxiv.org/abs/2407.21783}
\BIBentrySTDinterwordspacing

\bibitem{Llama3.2}
M.~L. 3.2,
  \url{https://ai.meta.com/blog/llama-3-2-connect-2024-vision-edge-mobile-devices/}
  [2024-01-17].

\bibitem{Vicuna}
\BIBentryALTinterwordspacing
L.~Zheng, W.-L. Chiang, Y.~Sheng, S.~Zhuang, Z.~Wu, Y.~Zhuang, and e.~a.
  Zi~Lin, ``Judging llm-as-a-judge with mt-bench and chatbot arena,'' 2023.
  [Online]. Available: \url{https://arxiv.org/abs/2306.05685}
\BIBentrySTDinterwordspacing

\bibitem{ShareGPT}
L.~Chen, J.~Li, X.~Dong, P.~Zhang, C.~He, J.~Wang, and e.~a. Zhao, Feng,
  ``Sharegpt4v: Improving large multi-modal models with better captions,'' in
  \emph{Proc. ECCV}.\hskip 1em plus 0.5em minus 0.4em\relax Springer, 2025, pp.
  370--387.

\bibitem{Phi-2}
M.~Javaheripi, S.~Bubeck, M.~Abdin, J.~Aneja, S.~Bubeck, C.~C.~T. Mendes, and
  e.~a. Chen, Weizhu, ``Phi-2: The surprising power of small language models,''
  \emph{Microsoft Research Blog}, vol.~1, no.~3, p.~3, 2023.

\bibitem{Phi-4-mini-instruct}
\BIBentryALTinterwordspacing
Microsoft, :, A.~Abouelenin, A.~Ashfaq, A.~Atkinson, H.~Awadalla, J.~B.
  Nguyen~Bach, and et~al., ``Phi-4-mini technical report: Compact yet powerful
  multimodal language models via mixture-of-loras,'' 2025. [Online]. Available:
  \url{https://arxiv.org/abs/2503.01743}
\BIBentrySTDinterwordspacing

\bibitem{Qwen2.5-Coder}
\BIBentryALTinterwordspacing
B.~Hui, J.~Yang, Z.~Cui, J.~Yang, D.~Liu, L.~Zhang, and e.~a. Tianyu~Liu,
  ``Qwen2.5-coder technical report,'' 2024. [Online]. Available:
  \url{https://arxiv.org/abs/2409.12186}
\BIBentrySTDinterwordspacing

\bibitem{frantar2023gptqaccurateposttrainingquantization}
\BIBentryALTinterwordspacing
E.~Frantar, S.~Ashkboos, T.~Hoefler, and D.~Alistarh, ``Gptq: Accurate
  post-training quantization for generative pre-trained transformers,'' 2023.
  [Online]. Available: \url{https://arxiv.org/abs/2210.17323}
\BIBentrySTDinterwordspacing

\bibitem{zhao2025arxiv}
\BIBentryALTinterwordspacing
J.~L. T. T. X. W. Y.~H. Wayne Xin~Zhao, Kun~Zhou and Y.~Min, ``A survey of
  large language models,'' 2025. [Online]. Available:
  \url{https://arxiv.org/abs/2303.18223}
\BIBentrySTDinterwordspacing

\bibitem{zheng2024arxiv}
\BIBentryALTinterwordspacing
Y.~Zheng, R.~Zhang, J.~Zhang, Y.~Ye, Z.~Luo, Z.~Feng, and Y.~Ma,
  ``Llamafactory: Unified efficient fine-tuning of 100+ language models,''
  2024. [Online]. Available: \url{https://arxiv.org/abs/2403.13372}
\BIBentrySTDinterwordspacing

\bibitem{jimenez2024ICLR}
\BIBentryALTinterwordspacing
C.~E. Jimenez, J.~Yang, A.~Wettig, S.~Yao, K.~Pei, O.~Press, and K.~R.
  Narasimhan, ``{SWE}-bench: Can language models resolve real-world github
  issues?'' in \emph{Proc. ICLR 2024}, 2024. [Online]. Available:
  \url{https://openreview.net/forum?id=VTF8yNQM66}
\BIBentrySTDinterwordspacing

\end{thebibliography}

\end{document}